\documentclass[12pt]{article}
\usepackage{graphicx,bm}

\def\k{{\bm k}}
\def\u{{\bm u}}
\def\v{{\bm v}}
\def\f{{\bm f}}
\def\x{{\bm x}}

\def\etal{{\it et al. }}

\begin{document}

\begin{center}{{\Large \bf Local transfer and spectra of a diffusive
field advected by large-scale incompressible flows}\\~\\
Chuong V. Tran\\
School of Mathematics and Statistics, University of St Andrews\\
St Andrews KY16 9SS, United Kingdom}
\end{center}
\date{\today}

\baselineskip=16pt

\centerline{\bf Abstract}

This study revisits the problem of advective transfer and spectra of a 
diffusive scalar field in large-scale incompressible flows in the presence 
of a (large-scale) source. By ``large-scale'' it is meant that the spectral 
support of the flows is confined to the wave-number region $k<k_d$, where 
$k_d$ is relatively small compared with the diffusion wave number 
$k_\kappa$. Such flows mediate couplings between neighbouring wave numbers 
within $k_d$ of each other only. It is found that the spectral rate of 
transport (flux) of scalar variance across a high wave number $k>k_d$ 
is bounded from above by $Uk_dk\Theta(k,t)$, where $U$ denotes the maximum 
fluid velocity and $\Theta(k,t)$ is the spectrum of the scalar variance, 
defined as its average over the shell $(k-k_d,k+k_d)$. For a given flux, 
say $\vartheta>0$, across $k>k_d$, this bound requires 
$$\Theta(k,t)\ge \frac{\vartheta}{Uk_d}k^{-1}.$$ This is consistent 
with recent numerical studies and with Batchelor's theory that predicts 
a $k^{-1}$ spectrum (with a slightly different proportionality constant) 
for the viscous-convective range, which could be identified with 
$(k_d,k_\kappa)$. Thus, Batchelor's formula for the 
variance spectrum is recovered by the present method in the form of a 
critical lower bound. The present result applies to a broad range of 
large-scale advection problems in space dimensions $\ge2$, including 
some filter models of turbulence, for which the turbulent velocity field
is advected by a smoothed version of itself. For this case, $\Theta(k,t)$
and $\vartheta$ are the kinetic energy spectrum and flux, respectively.

~

\centerline{xxxxxxxxxxxxxxxxxxxxxxxxx}

\section{Introduction}
The problem of scalar transport and mixing in turbulent fluid flows has 
been a subject of active 
research for decades, dating back to the late 1940s. Early studies by 
Obukhov \cite{Obukhov49} and Corrsin \cite{Corrsin51} applied Kolmogorov's 
theory of turbulence in a straightforward manner. They found that the 
scalar (fluid temperature in their case) variance behaved in the same 
manner as the turbulent kinetic energy, cascading via a $k^{-5/3}$ range
to a diffusion range at high wave numbers $k$ for disposal. This 
result is supposed to apply to cases of relatively small diffusivity 
$\kappa$ and viscosity $\nu$ in the regime $\kappa\approx\nu$, for which 
the viscous dissipation and diffusion ranges coincide. Batchelor 
\cite{Batchelor59} considered turbulent flows at moderate Reynolds numbers 
in the regime of large Prandtl or Schmidt number $P_r=\nu/\kappa\gg1$, 
for which there exists a broad viscous-convective range 
$k_\nu\ll k\ll k_\kappa$ 
between the viscous dissipation wave number $k_\nu$ and diffusion wave 
number $k_\kappa$. He found that in this range, the scalar variance 
spectrum $F(k)$ scales as $k^{-1}$ and is given~by
\begin{eqnarray}
\label{Bspectrum}
F(k)=\frac{\chi}{\gamma}\, k^{-1},
\end{eqnarray} 
where $\chi$ is the rate at which the scalar variance is dissipated,
i.e. the spectral rate of variance transport or variance flux, and $\gamma$ 
is an effective least-rate-of-strain parameter given by
$\gamma=C(\epsilon/\nu)^{1/2}$. Here $\epsilon$ denotes the mean rate of 
kinetic energy dissipation and $C$ is a constant of order unity.
From the Obukhov--Corrsin and Batchelor theories one may visualize a 
picture of scalar advection in flows at moderate Reynolds numbers in the 
limit of large $P_r$, in which a hybrid spectrum obeys the 
Obukhov--Corrsin $k^{-5/3}$ scaling in the fluid inertial range followed 
by the Batchelor $k^{-1}$ scaling in the viscous-convective range
\cite{Borgas04}. These pioneering theories have been considered to be 
breakthroughs and attracted considerable interest to the subject during 
its infancy [5-9]. Recently, fundamental issues in geophysical, 
environmental and industrial applications have sparked a surge in the 
area, resulting in a huge body of research [10-30] on a variety of 
dynamical aspects. Another reason for this surge is that
computers have become increasingly capable of taking on a scientific 
problem of this magnitude. Within the past few years, numerical evidence 
in support of the Batchelor theory and its predicted $k^{-1}$ spectrum 
has accumulated considerably \cite{Borgas04,Yeung02,Burton08}. However, 
this is far from conclusive as the viscous-convective ranges accessible 
to modern computers are still quite limited. Furthermore, a number of 
studies [16, 24-26] have either 
argued for or found spectra shallower than the Batchelor spectrum. For 
these reasons, as well as the phenomenological nature of the 
Obukhov--Corrsin and Batchelor theories, further theoretical consideration 
and numerical analysis (with higher resolutions whenever possible)
continue to be desirable.

In this study, we revisit the advection-diffusion problem, by carrying 
out a simple but rigorous analysis of the advective transfer term leading 
to a conclusion that is consistent with the Batchelor picture 
\cite{Batchelor59} and with recent numerical results 
\cite{Borgas04,Yeung02,Burton08}. We consider large-scale flows, meaning 
that the tail of the Fourier representation of the flows beyond some 
finite wave number $k_d$ is identically zero or at least can be ignored. 
Such smooth flows are relevant for practical purposes as most 
advection-diffusion problems in the geophysical and environmental
contexts are primarily concerned with large-scale advecting flows. They 
may even model Navier--Stokes turbulence at moderate Reynolds 
numbers if $k_d$ belongs to the viscous dissipation range and if the 
exponentially decaying tail of the velocity fields beyond $k_d$ can be 
ignored. These large-scale flows can mediate transfer of the scalar 
variance between neighbouring wave numbers within $k_d$ of each other 
only. It is found that the variance 
flux across a high wave number $k>k_d$ is bounded from above by 
$Uk_dk\Theta(k,t)$, where $U$ denotes the maximum fluid velocity and 
$\Theta(k,t)$ is the spectrum of the scalar variance, defined as its 
average over the shell $(k-k_d,k+k_d)$. From this upper bound, it can be
readily deduced that in the high $k$ limit, the flux diminishes if 
$\Theta(k,t)$ becomes steeper than $k^{-1}$. Hence, no spectra steeper 
than $k^{-1}$ could support a nonzero variance flux to the vicinity of 
the diffusion wave number $k_\kappa$ in the limit of large $k_\kappa/k_d$, 
which may be identified with $P_r$ (or $P_r^{1/2}$) \cite{T08}. Given a 
persistent source and in the limit $k_\kappa/k_d\to\infty$, this result
implies a divergence of the scalar variance not slower than logarithmic
in $k$ even if a variance cascade to the small scales is realizable. The 
present results apply to both passive and active scalars in large-scale 
flows, provided that in the active case, the excitation of the wave numbers
$k>k_d$ of the flows by nonlinear feedback mechanism can be ignored. They 
also apply to filter models of turbulence, for which the full turbulent 
velocity is advected by a smoothed version of itself. In this case,
$\Theta(k,t)$ is replaced by the kinetic energy spectrum, and the flux in
question is the kinetic energy flux.

\section{Preliminaries}

In this section, we briefly describe the advection-diffusion equations 
in spectral form, principally to illustrate the coupling locality, which
plays a key role in the present analysis. We then recall the variance 
conservation law and set out a few notations employed in this paper.

The advection-diffusion equations governing the evolution of a diffusive 
field $\theta(\x,t)$ advected by incompressible flows $\u(\x,t)$ are
\begin{eqnarray}
\label{governing}
\frac{\partial\theta}{\partial t} + \u\cdot\nabla\theta &=& 
\kappa\Delta\theta + f, ~~~~ \nabla\cdot\u = 0,
\end{eqnarray}
where $\kappa$ is the diffusivity and $f(\x,t)$ is a (large-scale) source. 
The spectral support of $\u(\x,t)$ is assumed to be confined to the
region $k<k_d$, where $k_d$ is a finite wave number. We consider Eq.\ 
(\ref{governing}) 
in an $n$-dimensional ($n\ge2$) periodic domain, enabling us to express 
our results conveniently in terms of spatial averages of dynamical 
quantities. These results can be seen to carry over to an unbounded domain 
with minimal change. All fields are assumed to have zero spatial average. 
The advected field $\theta(\x,t)$ can be either passive or active. In the 
latter case, the nonlinear feedback mechanism by $\theta(\x,t)$ on 
$\u(\x,t)$ can be arbitrary, as long as it does not ``irregularize'' 
$\u(\x,t)$ by exciting the small scales of $\u(\x,t)$ corresponding to 
$k>k_d$ to the extent that these scales can no longer be ignored. 
Furthermore, $\theta(\x,t)$ can be a vector, such as the fluid velocity 
in some filter models of turbulence (for which a pressure term is included).

The Fourier representations of $\u(\x,t)$ and $\theta(\x,t)$ are
\begin{eqnarray}
\label{fourier1}
\u(\x,t) &=& \sum_{k=|\k|<k_d}\widehat\u(\k,t)\exp\{i\k\cdot\x\}
\end{eqnarray}
and
\begin{eqnarray}
\label{fourier2}
\theta(\x,t) &=& \sum_\k\widehat\theta(\k,t)\exp\{i\k\cdot\x\},
\end{eqnarray}
respectively. Here $\k\neq0$ is the wave vector and $\widehat\u(\k,t)$ 
and $\widehat\theta(\k,t)$ are the Fourier transforms of $\u(\x,t)$ 
and $\theta(\x,t)$, respectively. The reality of $\u(\x,t)$ and 
$\theta(\x,t)$ requires $\widehat\u(\k,t)=\widehat\u^*(-\k,t)$ and 
$\widehat\theta(\k,t)=\widehat\theta^*(-\k,t)$. The incompressibility 
of $\u(\x,t)$ further requires $\k\cdot\widehat\u(\k,t) = 0$. In
spectral form, the first equation of Eq.\ (\ref{governing}) becomes
\begin{eqnarray}
\label{modal}
\frac{\partial}{\partial t}\widehat\theta(\k,t)
&=& \sum_{\k=\k'+\k''}\k'\cdot\widehat\u(\k'',t)\widehat\theta(\k',t) 
-\kappa k^2\widehat\theta(\k,t) + \widehat{f}(\k,t),
\end{eqnarray}
where $\widehat{f}(\k,t)$ is the Fourier transform of $f(\x,t)$.
The incompressibility of $\u(\x,t)$ manifests itself in Eq.\ (\ref{modal}) 
through the fact that $\k'\cdot\widehat\u(\k'',t)\widehat\theta(\k',t)=0$
if $\k'$ and $\k''$ are collinear. The triad relation $\k=\k'+\k''$, 
together with the constraint $k''=|\k''|<k_d$, implies that $k'=|\k'|$ 
satisfies $|k-k'|<k_d$. This means that a given $k>k_d$ can couple 
with other wave numbers within the shell $(k-k_d,k+k_d)$ only. This 
coupling locality has a significant consequence as will be seen in the
next section.

Given a ``reference'' wave number $k_*$ ($k_*\gg k_d$), let us denote by 
$\theta_<$, $\theta_>$, $\theta_i$, and $\theta_e$ the components of 
$\theta$ spectrally supported by the ball $b=\{\k:k\le k_*\}$, its 
complement $B=\{\k:k>k_*\}$, the inner shell $S_i=\{\k:k_*-k_d<k\le k_*\}$, 
and the outer shell $S_e=\{\k:k_*<k<k_*+k_d\}$, respectively (see Fig.\ 1). 
For example,
\begin{eqnarray}
\theta_>(\x,t)=\sum_{\k\in B}\widehat\theta(\k,t)\exp\{i\k\cdot\x\}.
\end{eqnarray}
For $k>k_d$, let $\Theta(k,t)$ denote the averaged variance taken 
over the shell $(k-k_d,k+k_d)$. For example, $\Theta(k_*,t)$ is given by
\begin{eqnarray}
\label{spectrum}
\Theta(k_*,t)&=&\frac{\langle\theta_i^2\rangle+\langle\theta_e^2\rangle}{2k_d},
\end{eqnarray}
where $\langle\cdot\rangle$ denotes a spatial average. It is evident that 
$\Theta(k,t)$ approximates the usual spectrum $F(k,t)$. In general, the 
approximation can become increasingly better for higher $k$ as the shell 
$(k-k_d,k+k_d)$ becomes thinner, in the sense that the ratio of the shell 
thickness $2k_d$ to its radius $k$ becomes smaller. For power-law spectra, 
$\Theta(k,t)$ actually approaches $F(k,t)$ in the limit
$k/k_d\to\infty$. For example, for the Batchelor spectrum given by 
Eq.\ (\ref{Bspectrum}), we have
\begin{eqnarray}
\Theta(k_*) &=& \frac{\chi}{2k_d\gamma}\int_{k_*-k_d}^{k_*+k_d}k^{-1}\,dk
=\frac{\chi}{2k_d\gamma}\ln\frac{k_*+k_d}{k_*-k_d}
=\frac{\chi}{2k_d\gamma}\ln\left(1+\frac{2k_d}{k_*-k_d}\right),
\end{eqnarray}
which indeed tends to $\chi(\gamma k_*)^{-1}$ in the limit 
$k_*/k_d\to\infty$. Thus $\Theta(k,t)$ tends to $F(k,t)$ in the limit 
of high $k$ (at least for power-law spectra). We will express our results 
in terms of $\Theta(k,t)$ instead of $F(k,t)$ since the former arises 
more naturally in the present context. 

Given periodic functions $\phi$ and $\psi$ having zero mean and
bounded mean-square gradients, we have
\begin{eqnarray}
\label{conservation}
\langle\phi\u\cdot\nabla\psi\rangle = - \langle\psi\u\cdot\nabla\phi\rangle.
\end{eqnarray}
This identity gives rise to a wealth of conservation laws, particularly
the variance conservation law, and is used repeatedly in this study. 

\section{Main results}

We now present the main results of this study. First, we elaborate on 
the locality of the variance transfer and then derive the lower bound for
$\Theta(k,t)$ as described earlier. Second, we show that given a 
persistent source, $\langle\theta^2\rangle$ grows without bound in the 
limit $k_\kappa\to\infty$, irrespective of the underpinning dynamics. 
Third, we compare the present finding with a recent result \cite{T08} 
derived on the assumption of bounded velocity gradients, i.e. 
$|\nabla\u|<\infty$, which is a weaker condition than the present 
one. Finally, the slight discrepancy between the present finding and 
Batchelor's formula (\ref{Bspectrum}) is discussed.

\subsection{Bounds for variance flux and spectrum}
The governing equation for the evolution of the small-scale variance 
$\langle\theta_>^2\rangle$ is obtained by multiplying Eq.\ (\ref{governing}) 
by $\theta_>$ and taking the spatial average of the resulting equation,
\begin{eqnarray}
\label{evolution1}
\frac{1}{2}\frac{d}{dt}\langle\theta_>^2\rangle 
+ \kappa\langle|\nabla\theta_>|^2\rangle
&=& -\langle\theta_>\u\cdot\nabla\theta\rangle
= -\langle\theta_>\u\cdot\nabla\theta_<\rangle,
\end{eqnarray}
where Eq.\ (\ref{conservation}) and the linearity of the advection term have 
been used and the forcing term vanishes
as $B$ is assumed to be source free. The triple-product term (flux term)  
in Eq.\ (\ref{evolution1}) represents the net variance transfer across $k_*$ 
into the region $k>k_*$, which drives the small-scale dynamics. At the 
modal level, this flux term consists of triple-product terms of the form 
$\widehat\theta(\k,t)\widehat\theta(\k',t)\k'\cdot\widehat\u(\k'',t)$, 
where $\k\in B$, $\k'\in b$ and $\k=\k'+\k''$. Since $k''<k_d$, this 
triad relation implies that $k$ and $k'$ can couple only if $k-k'<k_d$. 
Hence, only modes in $\theta_i$ and $\theta_e$, i.e. within the wave 
number shell $(k_*-k_d,k_*+k_d)$, contribute to the flux term 
(see Fig.\ 1). For this reason, the variance transfer can be considered 
as being highly local, particularly at high $k$, where the shell 
$(k_*-k_d,k_*+k_d)$ becomes relatively thin (radius becoming larger but 
thickness remaining fixed). Here, we use the term ``highly local'' to 
emphasize the fact that $k/k'\to1$ in the limit $k_*/k_d\to\infty$. This 
term is to distinguish the present couplings from those of a lesser 
degree of locality between $k\approx k'\approx k_*$ via $k''\approx k_*$, 
where the ratio $k/k'$ remains strictly greater than
unity in the same limit. Such couplings are clearly absent from the flux 
term. From this geometric consideration, we can write
\begin{eqnarray}
\label{id}
\langle\theta_>\u\cdot\nabla\theta_<\rangle =
\langle\theta_e\u\cdot\nabla\theta_i\rangle. 
\end{eqnarray}
Substituting this result into Eq.\ (\ref{evolution1}) yields
\begin{eqnarray}
\label{evolution2}
\frac{1}{2}\frac{d}{dt}\langle\theta_>^2\rangle 
+ \kappa\langle|\nabla\theta_>|^2\rangle
&=& -\langle\theta_e\u\cdot\nabla\theta_i\rangle
\le U\langle|\theta_e||\nabla\theta_i|\rangle 
\le U\langle|\theta_e|^2\rangle^{1/2}\langle|\nabla\theta_i|^2\rangle^{1/2}
\\
&\le& Uk_*\langle|\theta_e|^2\rangle^{1/2}\langle|\theta_i|^2\rangle^{1/2}
\le \frac{U k_*}{2}\left(\langle|\theta_e|^2\rangle 
+ \langle\theta_i^2\rangle\right)
= Uk_d k_*\Theta(k_*,t), \nonumber
\end{eqnarray}
where, as we recall, $U$ denotes the maximum fluid velocity and 
$\Theta(k,t)$ is the variance spectrum defined by Eq.\ (\ref{spectrum}).
In Eq.\ (\ref{evolution2}), the Cauchy--Schwarz inequality and the 
self-explanatory (Poincar\'e-type) inequality 
$\langle|\nabla\theta_i|^2\rangle\le k_*^2\langle\theta_i^2\rangle$
have been used. The bound for the flux term in Eq.\ (\ref{evolution2}) 
is interesting and can be readily interpreted in what follows.

\begin{figure}
\begin{center}
\includegraphics*[4cm,13.5cm][15cm,21cm]{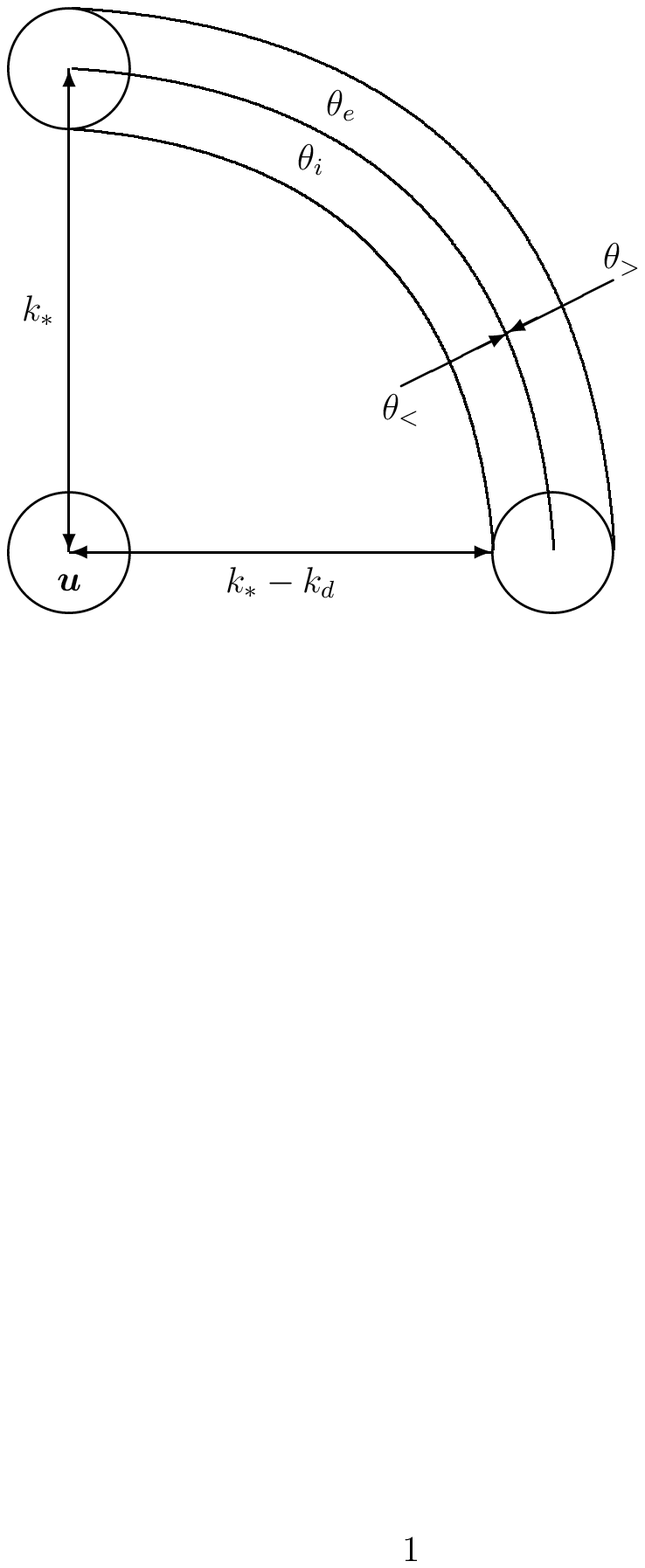}
\end{center}

F{\scriptsize IGURE}~1. A schematic description of the spectral supports 
for $\u$ and for the components $\theta_<,~\theta_>,~\theta_i,$ and
$\theta_e$ of $\theta$. The flux term involves only the wave numbers 
within the shell $(k_*-k_d,k_*+k_d)$, which supports $\theta_i$ and $\theta_e$.
      
\end{figure}

For a positive flux through $k_*$, say $\vartheta_*$, the final estimate 
in Eq.\ (\ref{evolution2}) implies that 
\begin{eqnarray}
\label{fluxbound}
\vartheta_* \le Uk_d k_*\Theta(k_*,t),
\end{eqnarray}
or equivalently
\begin{eqnarray}
\label{specbound}
\Theta(k_*,t) \ge \frac{\vartheta_*}{U k_d}k_*^{-1}.
\end{eqnarray}
It follows that a positive $k$-independent flux is possible only if 
$\Theta(k,t)$ becomes no steeper than $k^{-1}$ (pointwise) for high 
$k$. This constraint is consistent with Batchelor's theory that predicts 
a $k^{-1}$ spectrum for the viscous-convective range, which could be 
identified with $(k_d,k_\kappa)$. Since Eq.\ (\ref{specbound}) implies 
a divergence of $\langle\theta^2\rangle$ toward the small scales at
least as rapid as logarithmic in $k$, a positive variance flux to 
ever smaller scales (including those that diminish no more rapidly than
$(\ln k)^{-1}$) requires a priori an unbounded variance ``passage''.
This is in a sharp contrast to the classical direct energy cascade
(and the Obukhov--Corrsin variance cascade), which is supposed to 
proceed through an inertial range virtually free of energy. In some sense,
the energy cascade is rather ``rushing'', whereas the variance cascade of
the present case (if realizable) would be far less dramatic, ``leaking'' 
through a fully
filled inertial range. In the presence of a persistent scalar source, 
$\langle\theta^2\rangle$ necessarily  grows without bound in the limit 
$k_\kappa\to\infty$, for obvious reasons. On the one hand, a variance 
cascade to ever-smaller scales already requires at least a logarithmic 
divergence of $\langle\theta^2\rangle$ toward the small scales. On the 
other hand, if such a cascade is unrealizable, the injected variance 
is necessarily trapped at the large scales, thereby resulting in their 
unbounded growth. For the sake of completeness, this argument will be made 
more quantitative in the next subsection.

\subsection{Unbounded variance growth in the limit $k_\kappa\to\infty$}
Similar to Eq.\ (\ref{evolution1}), the governing equation for the evolution 
of the large-scale variance $\langle\theta_<^2\rangle$ is obtained by 
multiplying Eq.\ (\ref{governing}) by $\theta_<$ and taking the spatial 
average of the resulting equation,
\begin{eqnarray}
\label{evolution3}
\frac{1}{2}\frac{d}{dt}\langle\theta_<^2\rangle 
+ \kappa\langle|\nabla\theta_<|^2\rangle 
&=& -\langle\theta_<\u\cdot\nabla\theta\rangle + \vartheta
= \langle\theta_>\u\cdot\nabla\theta_<\rangle + \vartheta \nonumber\\
&=& \langle\theta_e\u\cdot\nabla\theta_i\rangle + \vartheta 
\ge -Uk_d k_*\Theta(k_*,t) + \vartheta,
\end{eqnarray}
where Eqs.\ (\ref{conservation}) and (\ref{id}) have been used and the 
inequality is a straightforward application of the upper bound for the 
flux term derived in Eq.\ (\ref{evolution2}). In Eq.\ (\ref{evolution3}), 
$\vartheta=\langle\theta_<f\rangle$ is the scalar variance injection 
rate. For some large time $t=T$, say $T=1/(2\kappa k_*^2)$, let 
$\overline{Q}$ denote the average over $[0,T]$ of a dynamical quantity 
$Q$. Taking the time average of Eq.\ (\ref{evolution3}) and re-arranging the 
terms in the resulting equation yields
\begin{eqnarray}
\label{average}
\kappa k_*^2\langle\theta_<^2\rangle +
\kappa\overline{\langle|\nabla\theta_<|^2\rangle}
&\ge&
\overline\vartheta - Uk_d k_*\overline{\Theta}(k_*),
\end{eqnarray}
where the initial value of $\langle\theta_<^2\rangle$ has been omitted 
for convenience. Upon making the substitution  
$\langle|\nabla\theta_<|^2\rangle\le k_*^2\langle\theta_<^2\rangle$
in Eq.\ (\ref{average}), we obtain
\begin{eqnarray}
\label{average1}
\kappa k_*^2\langle\theta_<^2\rangle + 
\kappa k_*^2\overline{\langle\theta_<^2\rangle}
&\ge&
\overline\vartheta - Uk_d k_*\overline{\Theta}(k_*).
\end{eqnarray}
In accord with a persistent source, let us assume $\overline\vartheta>0$. 
Now in the limit $k_\kappa\to\infty$ ($\kappa\to0$), if there exists no 
$k_*<\infty$ such that the right-hand side of Eq.\ (\ref{average1}) is 
positive, then $\langle\theta^2\rangle$ diverges toward the small scales 
as discussed above. On the other hand, if there exists $k_*<\infty$ such 
that the right-hand side of Eq.\ (\ref{average1}) is positive, then 
$(\langle\theta_<^2\rangle+\overline{\langle\theta_<^2\rangle})\to\infty$. 
It follows that $\langle\theta_<^2\rangle\to\infty$, and hence
$\langle\theta^2\rangle\to\infty$. Thus, $\langle\theta^2\rangle$ 
diverges regardless of whether or not there is a variance cascade.

\subsection{Discussion}
When $\u(\x,t)$ is not restricted to the large scales, there are no
constraints on $k'$ and $k$ in the triple-product terms 
$\widehat\theta(\k,t)\widehat\theta(\k',t)\k'\cdot\widehat\u(\k'',t)$
contributing to $\langle\theta_>\u\cdot\nabla\theta_<\rangle$. The flux 
term then involves, in principle, couplings for every $k'\le k_*$ and 
$k>k_*$. The presence of nonlocal couplings (between $k'\ll k_*$ 
and $k\gg k_*$ via $k''\approx k$) and the other type of local couplings 
(between $k\approx k'\approx k_*$ via $k''\approx k_*$) mentioned earlier 
effectively makes the flux term unmanageable by the present method, in 
the sense that its analytic estimates would be too excessive for 
meaningful interpretations. For this case, Tran \cite{T08} 
finds by a different approach that if the advecting velocity fields 
have bounded gradients, then diffusion anomaly, i.e. a variance cascade
to ever smaller scales, requires the variance at the small scales to 
be no less than that provided by the Batchelor $k^{-1}$ spectrum. This 
constraint is weaker than the present one as it does not rule out the 
possibility of bounded variance corresponding to non-power-law spectra 
having gaps of severe variance deficiency in the intermediate wave-number 
region, provided that the variance requirement at the small scales is 
met. The present finding, by exploiting the high locality of the variance 
transfer for large-scale advecting flows, rules out this possibility. 
The variance is required to grow without bound either via bounded 
spectra not steeper than $k^{-1}$ (pointwise) if a variance cascade 
is realizable or via unbounded spectra if otherwise. 

In the absence of a scalar source, a finite variance reservoir cannot
support a $k$-independent flux because such a flux requires an unbounded
variance ``passage'' as we have concluded. Our result allows for no 
significant ``chunk'' of a given initial variance reservoir 
$\langle\theta_0^2\rangle<\infty$ at large scales to break away and cascade
to the small scales by itself. Rather, it suggests a gradual spread out 
of $\langle\theta_0^2\rangle$ ever more thinly in wave-number space, 
giving rise to a diminishing flux, which can be readily estimated. 
Suppose that at a later time, a $k^{-1}$ range gets established from 
$k_d$ to $k_*\gg k_d$ or beyond. Then, in this range, the spectrum
$\Theta(k,t)$ is bounded by
$\Theta(k,t)\le\langle\theta_0^2\rangle k^{-1}/\ln(k_*/k_d)$. 
Upon substituting this into Eq.\ (\ref{fluxbound}), we obtain
\begin{eqnarray}
\label{fluxbound1}
\vartheta_* \le \frac{Uk_d\langle\theta_0^2\rangle}{\ln(k_*/k_d)}.
\end{eqnarray}
This means that $\vartheta_*$ diminishes at least as rapidly as 
$(\ln(k_*/k_d))^{-1}$. Note that although a logarithmic decay of the 
flux can be expected on heuristic grounds, Eq. (\ref{fluxbound1}) may 
not be rigorously derived without the constraint~(\ref{fluxbound}).

The present bound (\ref{specbound}) for $\Theta(k,t)$ resembles the 
Batchelor formula (\ref{Bspectrum}) in every aspect except that $Uk_d$ in 
Eq.\ (\ref{specbound}) plays the role of $\gamma$ in Eq.\ (\ref{Bspectrum}). 
This apparent discrepancy, however, can be reconciled if we reformulate 
the present problem in accord with the Batchelor setting. It can be seen 
that the product $Uk_d$ is essentially an upper bound for the velocity 
gradients $|\nabla\u|$. So if we identify 
$\nu(Uk_d)^2\approx\nu|\nabla\u|^2$ with the kinetic energy dissipation 
rate $\epsilon$ in the Batchelor setting of turbulent advection, then 
we obtain $Uk_d\approx (\epsilon/\nu)^{1/2}\approx\gamma$. Hence, Eqs.\ 
(\ref{Bspectrum}) and (\ref{specbound}) agree. This is no surprise 
because the Batchelor problem would reduce to the present case upon 
the hypothesis that the exponentially decaying tail (beyond $k_d$) of 
the turbulent velocity field contributes negligibly to the advective 
variance transfer. 

As it stands, Eq.\ (\ref{specbound}) captures the intuitive physical fact 
that for fixed $U$, flows at larger scales (smaller $k_d$) are poorer 
transporters as scalar spectra having larger spectral amplitudes, i.e. 
larger factors $\vartheta_*/(Uk_d)$, would be required to support the 
variance flux $\vartheta_*$ across $k_*$.

\section{Turtulent energy transfer by large-scale advection}

The above results apply to the energy transfer by large-scale advection 
in turbulence. Namely, the advection of the turbulent velocity by its 
large-scale component alone results in a contributing energy flux that 
vanishes at high $k$ if the energy spectrum becomes steeper than $k^{-1}$. 
For the Kolmogorov $k^{-5/3}$ spectrum, this means that the large-scale 
advection contributes negligibly to the direct energy transfer. On physical 
grounds, this is consistent with the expectation that the large scales,
while advecting the turbulent eddies, do not stretch them significantly. 
The remaining of this paper is devoted to detailed elaboration of this fact.

We begin by recalling the Navier--Stokes equations
\begin{eqnarray}
\frac{\partial\v}{\partial t} + (\v\cdot\nabla)\v + \nabla p &=& 
\nu\Delta\v + \f, ~~~~
\nabla\cdot\v = 0,
\end{eqnarray}
where $\v(\x,t)$ is the fluid velocity, $p(\x,t)$ is the pressure and 
$\f(\x,t)$ is a large-scale forcing. Let $\u$ be a large-scale component 
of $\v$, as defined by Eq.\ (\ref{fourier1}), and $\u'$ be its small-scale 
complement, i.e. $\v=\u+\u'$. Furthermore, let $\v_<,~\v_>,~\v_i,~\v_e$ 
and $V(k)$ be defined in the same ways as 
$\theta_<,~\theta_>,~\theta_i,~\theta_e$ and $\Theta(k)$, respectively. 
Note that $V(k)$ is approximately twice the usual kinetic energy spectrum 
and that all the components of $\v$ so defined are incompressible. 
Similar to Eq.\ (\ref{evolution2}), the evolution of the small-scale energy 
$\langle|\v_>|^2\rangle/2$ is 
governed by
\begin{eqnarray}
\label{e-evolution}
\frac{1}{2}\frac{d}{dt}\langle|\v_>|^2\rangle 
+ \nu\langle|\nabla\v_>|^2\rangle 
&=& -\langle\v_>\cdot(\v\cdot\nabla)\v_<\rangle
= -\langle\v_>\cdot(\u\cdot\nabla)\v_<\rangle
-\langle\v_>\cdot(\u'\cdot\nabla)\v_<\rangle \nonumber\\
&=&
-\langle\v_e\cdot(\u\cdot\nabla)\v_i\rangle
-\langle\v_>\cdot(\u'\cdot\nabla)\v_<\rangle \nonumber\\
&\le& 
Uk_d k_*V(k_*,t) -\langle\v_>\cdot(\u'\cdot\nabla)\v_<\rangle, 
\end{eqnarray}
where the forcing and pressure terms vanish as the region under 
consideration is assumed to be force free and $\v_>$ is incompressible. 
In the final equation of Eq.\ (\ref{e-evolution}), the first term on the 
right-hand side represents an upper bound for the energy transfer across 
$k_*$ due to large-scale advection and the second term is the energy 
transfer across $k_*$ due to small-scale advection. The former vanishes 
for high $k_*$ if $V(k,t)$ becomes steeper than $k^{-1}$. This means 
that the latter is solely responsible for the direct energy cascade in 
the classical picture of turbulence, for which the $k^{-5/3}$ energy
inertial range is far too steep for the former to make a non-negligible 
contribution. At the modal level, this result is consistent with the 
expectation that triad interactions involving well-separated scales 
(those due to large-scale advection in the flux term 
$\langle\v_e\cdot(\u\cdot\nabla)\v_i\rangle$) are relatively weak. Note
that not all triads of well-separated scales are contained within
$\langle\v_e\cdot(\u\cdot\nabla)\v_i\rangle$ as the term 
$\langle\v_>\cdot(\u'\cdot\nabla)\v_<\rangle$ also has this type of
triads. Such triads are formed by large-scale modes in $v_<$ and 
small-scale modes in $\v_>$ and $\u'$. Like their counterparts in 
$\langle\v_e\cdot(\u\cdot\nabla)\v_i\rangle$, these can be shown to be 
relatively weak and not responsible for the classical direct energy cascade.

The above result may be applicable to models of turbulence that are
derived by regularizing the Navier--Stokes equations by a variety of 
filtering techniques (see Graham \etal \cite{Graham08} for a discussion 
of several such models). For example, let us consider the Leray model, 
obtained by dropping the small-scale component $\u'$ from the advecting
velocity in the Navier--Stokes system, i.e.
\begin{eqnarray}
\label{filter}
\frac{\partial\v}{\partial t} + (\u\cdot\nabla)\v + \nabla p &=& 
\nu\Delta\v + \f, ~~~~
\nabla\cdot\v = 0.
\end{eqnarray}
For this simple model, the governing equation for 
$\langle|\v_>|^2\rangle/2$ is given by Eq.\ (\ref{e-evolution}) without the 
small-scale advection term $\langle\v_>\cdot(\u'\cdot\nabla)\v_<\rangle$.
As a consequence, the classical direct energy cascade is not realizable 
for the reason discussed in the preceding paragraph. Instead, the energy 
behaves in the same manner as the variance $\langle\theta^2\rangle$ 
described earlier. Namely, the energy either cascades to the small scales 
via spectra not steeper than $k^{-1}$ or else accumulates at the large 
scales. Given a persistent source of energy, i.e. 
$\overline{\langle\v\cdot\f\rangle}>0$, the energy necessarily grows 
without bound in the inviscid limit. Equation (\ref{filter}) resembles 
regularization models of turbulence, which have been studied widely as 
alternatives to subgrid-scale models \cite{Graham08} and for which the 
$k^{-1}$ scaling for the energy spectrum has been found by 
phenomenological arguments. The present result provides a different 
perspective to this possible scaling.

In passing, we would like to note that the question of realizability of 
a (variance or energy) cascade and the associated $k^{-1}$ (or shallower) 
spectrum cannot be resolved by the present analysis. This question is
challenging because a lower bound for the flux term is highly infeasible,
even for very simple flows. Given this difficulty, one may be better off 
resorting to numerical methods. What we have shown here is that if there 
is a cascade, then it must proceed through spectra not steeper than 
$k^{-1}$ (pointwise). The critical $k^{-1}$ scaling can be seen as most 
plausible for a number of reasons. In particular, it would correspond to 
a cascade of maximal spectral extent. 

\section{Concluding remarks}

In summary, we have examined the advective transfer and spectral scaling 
of a diffusive field $\theta(\x,t)$ in large-scale incompressible flows 
$\u(\x,t)$, whose spectral support is confined to the wave-number region 
$k<k_d$, for some finite wave number $k_d$, which is relatively small as 
compared with the diffusion wavenumber $k_\kappa$. The main result obtained 
is the upper bound $Uk_dk_*\Theta(k_*,t)$ for the variance flux across a 
high wave number $k_*>k_d$. Here $U$ denotes the maximum fluid velocity 
and $\Theta(k,t)$ is the variance spectrum, defined as its average over 
the shell $(k-k_d,k+k_d)$. The derivation of this bound exploits the very 
fact that the advecting flows under consideration mediate variance transfer 
between neighbouring wave numbers within $k_d$ of each other only. The 
derived bound implies that for $k\gg k_d$, a nonzero $k$-independent flux 
is possible only if $\Theta(k,t)$ becomes no steeper than $k^{-1}$ 
(pointwise). This result is consistent with Batchelor's theory and with 
recent numerical and theoretical results [4, 22, 30, 31]. A novel element 
of the present findings is the pointwise constraint on $\Theta(k,t)$ in 
Eq.\ (\ref{specbound}). 
Given this constraint and a persistent source, the variance is required 
to grow without bound in the limit $k_\kappa/k_d\to\infty$ ($\kappa\to0$), 
irrespective of the underpinning dynamics. The present results have been
shown to apply to the Leray model of turbulence, for which the turbulent 
velocity is advected by a smoothed version of itself. Furthermore, they 
apply equally to different space dimensions $\ge2$ as the analysis is 
dimension-independent. Finally, the possible relation between the 
advecting and advected fields does not enter the calculations. Hence, 
the results are valid for both passive and active scalars, provided 
that in the latter case, the nonlinear feedback mechanisms on the flows 
do not alter their large-scale designation.

\end{document}